\documentclass[copyright,creativecommons]{eptcs}
\providecommand{\event}{TARK XVII} 
\usepackage{amsmath}
\usepackage{amssymb}
\usepackage{mathrsfs}
\usepackage{enumerate}

\title{Measuring Belief and Risk Attitude}
\author{Sven Neth
\institute{University of California, Berkeley}
\email{nethsven@berkeley.edu}}

\newcounter{first-gamble}

\def\titlerunning{Measuring Belief and Risk Attitude}
\def\authorrunning{Sven Neth}

\begin{document}

\maketitle

\begin{abstract}
Ramsey \cite{Ramsey1926} sketches a proposal for measuring the subjective probabilities of an agent by their observable preferences, assuming that the agent is an expected utility maximizer. I show how to extend the spirit of Ramsey's method to a strictly wider class of agents: \emph{risk-weighted expected utility maximizers}, introduced by Buchak \cite{Buchak2013}. In particular, I show how we can measure the \emph{risk attitudes} of an agent by their observable preferences, assuming that the agent is a risk-weighted expected utility maximizer. Further, we can leverage this method to measure the subjective probabilities of a risk-weighted expected utility maximizer. 
\end{abstract}

\section{Introduction}

How do we infer what people believe and desire from their observable behavior? One way to address this problem is to invoke some version of the \emph{principle of charity}: We attribute beliefs and desires to an agent that \emph{rationalize} the behavior we observe \cite{Davidson1973, Lewis1974}. 

What does it mean to be rational? One way to articulate what it means to be rational is the framework of \emph{decision theory}, a mathematical theory which defines optimal decisions given an subjective probability function and a utility function \cite{Savage1954, Jeffrey1990}. Ramsey \cite{Ramsey1926} shows how we can use the assumption that people follow one particular decision theoretic principle, viz. expected utility maximization, to infer their subjective probability function and utility function from their observable betting behavior.

In this paper, I focus on Ramsey's method to infer subjective probabilities from betting behavior after the utility function is already determined. Ramsey advises us to think of subjective probabilities as betting odds. However, when agents are sensitive to risk, subjective probabilities and betting odds can come apart, so Ramsey's method no longer works. I show that we can extend Ramsey's method, or something very similar to it, to agents which follow a weaker decision theoretic principle: risk-weighted expected utility maximization, defended by Buchak \cite{Buchak2013}.

\section{Measuring Subjective Probability}

Before explaining Ramsey's method for measuring subjective probability, let me briefly explain expected utility theory in the framework of Savage \cite{Savage1954}. 

\subsection{Expected Utility Theory}

We start with a sets of \emph{states} $\mathscr{S}$, which contains all the maximally specific ways the world might be. Subsets of $\mathscr{S}$ are called \emph{events} and form the event space $\mathscr{E}$, which is a $\sigma$-algebra on $\mathscr{S}$. We further have a set $\mathscr{O}$ of all possible \emph{outcomes}, which are the bearers of value for our agent. Following Savage, we are assuming \emph{state-independent utilities}, which means that the value of outcomes does not depend on the state of the world.

We model \emph{gambles} as functions from states to outcomes, that is, as functions from $\mathscr{S}$ to $\mathscr{O}$. (Savage calls these `acts'.) We assume that gambles assign different outcomes to only finitely many events. This means that we can write a gamble $g$ in the following form: $\{o_1, E_1;...;o_n, E_n\}$. This gamble yields outcome $o_1$ if event $E_1$ obtains, outcome $o_2$ if event $E_2$ obtains and so on. Further, $E_1, E_2,... E_n$ form a \emph{partition} of our state space $\mathscr{S}$, which means that they are mutually exclusive and collectively exhaustive.

We assume that our agent comes with a \emph{subjective probability function} $p$ on the event space $\mathscr{E}$, which models our agents' credences. Further, our agent comes with a \emph{utility function} $u$ on the outcome space $\mathscr{O}$, which models our agents' preferences over outcomes. Given this setup, the \emph{expected utility} (EU) of gamble $g$ is defined as follows: 
\begin{equation*}
EU(g) = \sum_{i =1}^{n} p(E_i)u(o_i).
\end{equation*}
Our agent has preferences over gambles, which are reflected in their observable choice behavior. We write $f \preceq g$ to mean that our agent prefers gamble $g$ to gamble $f$, and $f \sim g$ to mean that our agent is indifferent between gambles $f$ and $g$.\footnote{As usual, we define $f \sim g$ as $f \preceq g \land g \preceq f$ and $f \prec g$ as $f \preceq g \land g \not \preceq f$.} Our agents' preference ordering over gambles is \emph{representable} as EU maximization just in case, for any gambles $f$ and $g$, we have 
\begin{equation*}
f \preceq g \iff EU(f) \leq EU(g)
\end{equation*}
for some subjective probability function $p$ and utility function $u$. There are various \emph{representation theorems} for EU theory, which show that any preference ordering among gambles which satisfies certain conditions is representable as EU maximization relative to a unique probability function $p$ and a utility function $u$ unique up to positive linear transformation \cite{Savage1954}.

One common way to interpret Ramsey's posthumously published essay \emph{Truth and Probability} is as an (early and perhaps incomplete) representation theorem for EU theory \cite{Fishburn1981, Elliott2017}. I want to focus on a different aspect of Ramsey's essay, which comes out in the following passage:
\begin{quote}
``The subject of our inquiry is the logic of partial belief, and I do not think we can carry it far unless we have at least an approximate notion of what partial belief is, and how, if at all, it can be measured [...] We must therefore try to develop a purely psychological method of measuring belief.'' \cite[p. 166]{Ramsey1926}
\end{quote}
Ramsey is ultimately interested in \emph{measuring} the partial beliefs, or subjective probabilities, of an agent. To achieve this end, Ramsey suggests assuming that the agent is an EU maximizer:
\begin{quote}
``I suggest that we introduce as a law of psychology that his
behaviour is governed by what is called the mathematical expectation [...]'' \cite[p. 174]{Ramsey1926}
\end{quote} 
Ramsey goes on to list various conditions which an agents' preferences over gambles must satisfy in order to be representable as EU maximization. However, this representation theorem is merely a means to the end of measuring subjective probability.

\subsection{Ramsey's Method}

Let me now turn to explain the method for measuring subjective probability sketched by Ramsey \cite{Ramsey1926}.\footnote{Also see \cite{Savage1971, Bradley2004,  Bermudez2011}.} Assuming

\begin{enumerate}

\item  an agent's preference ranking among gambles is representable as EU maximization,  

\item we have already determined our agent's utility function $u$ (up to positive linear transformation),

\end{enumerate} 

Ramsey gives us a way to measure the subjective probability our agent assigns to an arbitrary event by observing our agents' preferences over gambles.\footnote{Note that Ramsey also sketches a method for measuring the utilities of our agent, but the details of this will not concern us here. The basic idea is that Ramsey tells us under which condition the utility difference between outcome $a$ and $b$ equals the utility difference between outcome $c$ and $d$, which fixes the utility function up to positive linear transformation \cite{Jeffrey1990, Parmigiani2009}.}

Ramsey's basic idea is that subjective probabilities are betting odds:
\begin{quote}
``This amounts roughly to defining the degree of belief in $p$ by the odds at which the subject would bet on $p$ [...]'' \cite[pp. 179-80]{Ramsey1926}
\end{quote}
Let me now explain how Ramsey's method works in more detail. We want to find out which subjective probability our agent ascribes to event $E$. First, we have to find three outcomes $b,m,w$ which satisfy what I will call \emph{Ramsey's conditions}: 

\begin{enumerate}[i.]

\item our agent strictly prefers $b$ over $w$: $w \prec b$,\footnote{Note that every outcome $o$ corresponds to a \emph{constant act}: a gamble which always yields outcome $o$ no matter what. To simplify notation, I will denote the constant act yielding $o$ by $o$.}

\item our agent is indifferent between getting $m$ for certain and a gamble which yields $b$ if $E$ happens and $w$ otherwise: $m \sim \{b, E; w, \textrm{not-}E\}$.

\end{enumerate}

Note that we have to assume that our agent's outcome space is sufficiently rich to contain such outcomes for any event $E$ whose probability we want to measure. By our first assumption, our agent maximizes EU, so
\begin{equation*}
u(m) = u(w) + p(E)(u(b) - u(w)),
\end{equation*}
which means that $p(E) = \frac{u(m) - u(w)}{u(b) - u(w)}$. Now, we can use our second assumption, that we have already determined our agent's utility function $u$, to compute $p(E)$. Since $E$ was an arbitrary event, Ramsey has shown us how to measure the subjective probability our agent assigns to arbitrary events.

To illustrate Ramsey's method, consider Alice, who only cares about money and values money linearly, so we have $u(\$x) = x$. Ramsey tells us to measure the subjective probability Alice ascribes to event $E$ by finding out for which price $x$ between zero and one dollar Alice is indifferent between getting $x$ cents for certain and a bet which pays one dollar if $E$ happens and nothing otherwise. Now we can use the assumption that Alice maximizes EU to infer that $p(E) = x$.\footnote{We have $p(E) = \frac{u(x) - u(0)}{u(1) - u(0)}$, so by the assumption that Alice values money linearly, $p(E) = \frac{x}{1} = x$.}

Note that Ramsey's method falls short of an \emph{algorithm} for determining the subjective probability an agent assigns to arbitary events. This is because Ramsey hasn't given us a general method for finding outcomes $b,m,w$ which satisfy Ramsey's conditions. Still, if we can always find these outcomes, we have a completely general way to determine the subjective probabilities of our agent by their observable betting behavior. 

\section{Beyond Expected Utility}

In this paper, I show that we can generalize Ramsey's method to agents who do not maximize expected utility. In particular, I generalize Ramsey's method to \emph{risk-weighted expected utility maximizers} by Buchak \cite{Buchak2013}. 

\subsection{Limits of Expected Utility Theory}

It is well known that people sometimes have preferences which are difficult to reconcile with expected utility theory. Here is an example first discussed by Allais \cite{Allais1953}:

\begin{quote}
\textbf{Allais paradox.}

You get to choose between the following two gambles:

\begin{enumerate}

\item One million dollars for certain.\label{L1}

\item 89 \% chance of winning one million dollars, 10 \% chance of winning five million dollars, 1 \% chance of winning nothing.\label{L2}
\setcounter{first-gamble}{\value{enumi}}
\end{enumerate}

\noindent Now, you get to choose between the following two gambles:

\begin{enumerate}
\setcounter{enumi}{\value{first-gamble}}
\item 89 \% chance of winning nothing, 11 \% chance of winning one million dollars.\label{L3}

\item 90 \% chance of winning nothing, 10 \% chance of winning five million dollars.\label{L4}

\end{enumerate}

\end{quote}

If you picked (\ref{L1}) and (\ref{L4}), you might be surprised to learn that these preferences are inconsistent with EU maximization. There is no way to assign utility values to dollar amounts that makes the expected utility of (\ref{L1}) higher than the expected utility of (\ref{L2}), while also making the expected utility of (\ref{L4}) higher than the expected utility of (\ref{L3}). Now as it turns out, many subjects exhibit just these preferences: they prefer (\ref{L1}) over (\ref{L2})  and (\ref{L4}) over (\ref{L3}) \cite{Machina1987, Oliver2009}. Call these the \emph{Allais-preferences}. 
 
There is a substantive question about whether it is \emph{rational} to have Allais-preferences. However, we do not need to decide this question here. What matters for our purposes is that real-life agents sometimes \emph{have} Allais-preferences, whether they are rational or not. This means that real-life agents are not always EU maximizers. Therefore, Ramsey's method is not always applicable to real-life agents.

\subsection{Risk-Weighted Expected Utility Theory}

There are alternative decision theories which are designed to accommodate Allais-preferences, such as \emph{risk-weighted expected utility theory} defended by Buchak \cite{Buchak2013}.\footnote{See \cite{Buchak2017} for an accessible introduction. REU theory builds on earlier work on rank-dependent
utility theory, in particular \cite{Quiggin1983, machina_more_1992, KW2003}.} In this theory, agents are characterized by a subjective probability function, a utility function and a \emph{risk function} $r$, which measures our agent's attitude towards risk. They do not maximize expected utility, but \emph{risk-weighted expected utility}. 

Given subjective probability function $p$, utility function $u$ and risk function $r$, the \emph{risk-weighted expected utility (REU)} of gamble $g = \{o_1,E_1;...;o_n, E_n\}$ with $u(o_1) \leq ... \leq u(o_n)$ is defined as follows:\footnote{We stipulate that $u(o_0) = 0$.}
\begin{equation*}
REU(g) = \sum\limits_{j=1}^n \left( r \left(\sum\limits_{i=j}^n p(E_i)\right) \left( u(o_j) - u(o_{j-1}) \right) \right).
\end{equation*}
To understand the difference between EU theory and REU theory, it is best to look at an example. Meet Alice and Bob. Both Alice and Bob value money linearly, so we have $u(\$x) = x$ for both Alice and Bob. Now suppose I have a fair coin and offer both Alice and Bob a gamble which yields one dollar if the coin comes up heads and nothing otherwise. Since both believe that the coin is fair, their subjective probability that it will come up heads is $\frac{1}{2}$. However, depending on their attitude towards risk, Alice and Bob might give more or less \emph{weight} to this event when deciding how much to pay for my gamble. Alice is indifferent towards risk, so Alice is willing to pay up to 50 cent for my gamble. In contrast, Bob prefers 50 cents for certain over my gamble. In fact, Bob is only willing to pay up to 10 cents for my gamble. Therefore, Bob can't be an EU maximizer. Rather, Bob is \emph{risk-averse}: for Bob, worse outcomes play a larger role in determining the value of a gamble than better ones.\footnote{Note that we assumed that Bob's utility function is a linear function of money. We could try to model risk averse agents within EU theory by concave utility functions. However, this approach faces several problems. First, there are risk-averse preferences which can't be captured in EU theory, such as the Allais paradox discussed above. Second, concave utility functions for small stakes predict extreme risk aversion for bigger stakes, see \cite{Rabin2000}. Further see \cite{Okasha2007} for an evolutionary explanation of why real-world agents are risk-averse in ways which are incompatible with EU theory.}

We can model the disagreement between Alice and Bob in the framework of REU theory. According to REU theory, two agents with the same probability and utility function can disagree about the value of a gamble if they have different attitudes towards risk. We can write my gamble from above as $f =\{\$1, H ; \$0,T\}$, which yields one dollar if the coin lands heads and zero dollar otherwise. The risk-weighted expected utility of $f$ is
\begin{equation*}
REU(f) = u(\$0) + r(p(H))((u(\$1) - u(\$0)) = r(p(H)).
\end{equation*}
Intuitively, $r$ measures how much \emph{weight} our agent puts on the event that the coin lands heads.  An agent who is indifferent towards risk, such as Alice, weights each event with its subjective probability. (This means that EU maximization is the special case of REU maximization where $r$ is the identity function.) A risk averse agent, such as Bob, weights an event with \emph{less} than its subjective probability if the event is not certain. A risk loving agent weights an event with \emph{more} than its subjective probability if the event is not certain. Thus, the risk function maps subjective probability to \emph{decision weight}, where the decision weight of an event is the contribution which the probability of the event makes to the value of a gamble in which this event yields the best outcome. 

Following Buchak \cite{Buchak2013}, I assume that the risk function $r$ satisfies the following constraints:
\begin{enumerate}[i.]
\item $r: [0,1] \to [0,1]$,
\item $r(1) = 1$ and $r(0) = 0$,
\item $r$ is strictly increasing: $a < b$ implies $r(a) < r(b)$,
\item $r$ is continuous,
\end{enumerate}  
all of which correspond to natural constraints on the relationship between subjective probability and decision weight. Both subjective probability and decision weights are numbers between zero and one, the decision weight of a certain event should be maximal and the decision weight of an impossible event minimal, more probable events should get higher decision weights and the decision weight of an event should vary continuously with its subjective probability.

If you think that it is rational to be sensitive towards risk in the way modeled by REU theory, this is a good reason to think that REU theory, and not EU theory, is the correct normative theory of decision making. However, I will not defend this claim here. I am merely committed to the weaker claim that REU theory is more adequate than EU theory when it comes to \emph{describing} and \emph{interpreting} the behavior of real-word agents.\footnote{See \cite{Buchak2016} for a discussion of the difference between normative, descriptive and interpretative uses of decision theory.} Like Ramsey, I am interested in REU theory as a `general psychological theory'.

\subsection{Limits of Ramsey's Method}

Given that the goal of Ramsey is to measure to beliefs of real-world agents, it would be great if we could use Ramsey's method to measure the subjective probabilities of REU maximizers. However, as it stands, Ramsey's method does not work for REU maximizers. 

Our example above shows that while Ramsey's method works for agents like Alice, who are indifferent towards risk, it does not work for risk averse agents like Bob. According to Ramsey's method, the subjective probability that Bob ascribes to the event that the coin comes up heads is .1, but that is not the right prediction. From the perspective of REU theory, the problem is that Ramsey's method conflates subjective probabilities and betting odds, which are the \emph{joint product} of subjective probability and attitude towards risk. If you are sensitive to risk, then your betting odds are not identical to your subjective probabilities, so measuring your subjective probabilities by looking at your betting odds is a bad idea.

We can make this point a bit more precise. Suppose we want to use Ramsey's method to measure which subjective probability a REU maximizer ascribes to event $E$. We first find outcomes $b,m,w$ which satisfy Ramsey's conditions. Then, we use the assumption that our agent maximizes REU to infer that
\begin{equation*}
u(m) = u(w) + r(p(E))(u(b) - u(w)),
\end{equation*}
so $r(p(E)) = \frac{u(m) - u(w)}{u(b) - u(w)}$. However, now we are stuck. We know $r(p(E))$, the decision weight of event $E$, which is the joint product of subjective probability and attitude towards risk. Note that $r(p(E))$ is not equal to $p(E)$, unless we are considering an agent who is indifferent towards risk. So the question is: \emph{How can we separate the distinct contributions of risk attitude and subjective probability?}

In the rest of this paper, I sketch an answer to this question. The core idea of my proposal is a method to measure the risk attitude of a REU maximizer by their observable preferences among gambles, similar to Ramsey's method to measure the subjective probabilities of an EU maximizer by their observable preferences among gambles. 

Note that my project is different from a representation theorem for REU theory. Buchak \cite{Buchak2013} proves a representation theorem for REU theory, which shows that any preference ranking on gambles which satisfy certain conditions can be represented as REU maximization relative to a unique probability function $p$, a unique risk function $r$ and a utility function $u$ which is unique up to positive linear transformation. However, this representation theorem does not tell us how to \emph{construct} the risk function and probability function of a REU maximizer from their observable behavior. This is my goal in this paper. I am assuming that an agent is a REU maximizer and then use this fact to `reverse engineer' their risk function and probability function from their observable behavior.

\section{Measuring Risk Attitudes}

I now turn to explain how we can measure the risk function of an REU maximizer. I end by explaining how we can leverage this method to measure the subjective probabilities of a REU maximizer.

\subsection{Ramsey Meets Risk}

My goal is the following. Assuming 
\begin{enumerate}
\item  an agent's preference ranking among gambles is representable as REU maximization,  
\item we already determined our agent's utility function $u$ (up to positive linear transformation),\footnote{Note that we can determine the utility function of a REU maximizer in a similar way to Ramsey's approach to determine the utility function of a EU maximizer. We can write down a certain condition on gambles involving the outcomes $a,b,c,d$, viz. comonotonic tradeoff consistency, which ensures that $u(a) - u(b) = u(c) - u(d)$. See \cite[p. 103]{Buchak2013}.}
\end{enumerate} 
I describe a method to measure our agent's attitude towards risk by uniquely determining their risk function $r$. 

As explained above, the risk function maps subjective probabilities to decision weights. Now we already have a method for measuring the decision weight of an arbitrary event. As explained above, this is what Ramsey's original method does in the context of REU theory. So to measure our agent's risk function, we need some independent way to find events with known subjective probabilities. Then, we can use Ramsey's method to measure the decision weight of these events, and so infer the mapping from subjective probabilities to decision weights.

It is important to observe that to determine our agent's risk function $r$ uniquely, it suffices to determine the decision weights of all events with \emph{rational-valued} probabilities, that is, to determine $r(a)$ for all $a \in [0,1]\cap \mathbb{Q}$. This is because the risk function $r$ is continuous, and any real-valued continuous function is uniquely determined by its value on all rational points.\footnote{This, in turn, is because the rationals are dense in the reals: There is a rational number between every two distinct real numbers \cite[p. 20]{Abott2001}.} This reduces our problem to the following: We need an independent way to find events with known rational-valued subjective probabilities. Then, we can measure the decision weights of those events by Ramsey's method, and so determine $r(a)$ for all $a \in [0,1]\cap \mathbb{Q}$, and so determine $r$ uniquely.

\subsection{The Method of Fair Lotteries}

Our goal is to find events with known rational-valued subjective probability. To achieve this goal, we look back at Ramsey, who shows us how to find an event with subjective probability $\frac{1}{2}$.\footnote{See \cite[p. 19]{Ramsey1926}. This is \emph{en route} to Ramsey's derivation of utilities, which we don't focus on here.} First, we find two outcomes $b$ and $w$ such that our agent strictly prefers $b$ over $w$. Then, we find an event $E$ which satisfies the following condition:
\begin{equation*}
\{E, b ; \textrm{not-}E, w\} \sim  \{E, w ; \textrm{not-}E, b\},
\end{equation*}
so our agent is indifferent between a gamble which yields $b$ if $E$ happens and $w$ otherwise and another gamble which yields $b$ if $E$ doesn't happen and $w$ otherwise. This means that our agent doesn't care whether the `good prize' is on $E$ or not-$E$. Ramsey then uses the assumption that the agent is an EU maximizer to infer that $p(E) = \frac{1}{2}$. 

For our purposes, the crucial observation is that the assumption of  EU maximization is not essential to this argument. It also goes through with on weaker assumption that the agent is a REU maximizer. In fact, the assumption required to make the argument work is even weaker:
\begin{quote}
\textbf{Better Prize Condition.} If an agent strictly prefers $b$ to $w$ and is indifferent between the gamble $\{E, b ; \textrm{not-}E, w\}$ and the gamble  $\{E, w; \textrm{not-}E, b\}$, then our agent thinks that $E$ and not-$E$ are equally probable: $p(E) = p(\textrm{not-}E)$.
\end{quote}
This assumption is entailed by both EU maximization and REU maximization. The important observation is that while decision weights are not the same as probabilities, the fact that two events have \emph{equal} decision weight implies that they have \emph{equal} probability.\footnote{Here is the argument: Since $\{E, b ; \textrm{not-}E, w\} \sim  \{E, b; \textrm{not-}E, w \}$, we have $u(w) +r(p(E))(u(b) - u(w)) = u(w) +r(p(\textrm{not-}E))(u(b) - u(w))$, so $r(p(E)) = r(p(\textrm{not-}E))$.  Since $r$ is strictly increasing, and so injective, $p(E) = p(\textrm{not-}E)$.} Therefore, if an agent does not care whether the `good prize' is on event $E$ or not-$E$, our agent must think that the event $E$ and the event not-$E$ are equally probable, even if our agent is sensitive to risk. Intuitively, this is because the choice is between two gambles -- there are no certainties to be had. So whether or not you are sensitive to risk, you should prefer the gamble which has the better prize on the more probable event. Further, the only way for $E$ and not-$E$ to be equally probable is if both events have probability $\frac{1}{2}$. Therefore, we can use Ramsey's idea to find an event with subjective probability $\frac{1}{2}$.

Now the crucial point is the following: We can generalize Ramsey's idea to all rational numbers, using what I call the \emph{method of fair lotteries}. Pick any rational number $a$ in the $[0,1]$ interval, so $a \in  [0,1]\cap \mathbb{Q}$. We want to find some event $E$ with $p(E) = a$. Because $a$ is a rational number, we have $a = \frac{k}{n}$ for some natural numbers $k$ and $n$, where $k \leq n$. We can think of an event with probability $\frac{k}{n}$ in the following way: It is the event that one of the first $k$ tickets in a fair lottery with $n$ tickets in total wins. So to solve our problem, we have to find a lottery with $n$ tickets which our agent considers to be fair.

We can model a fair lottery with $n$ tickets as a partition $\{E_1, ... ,E_n\}$  of our state space $\mathscr{S}$ into $n$ equiprobable events. So, given a partition with $n$ events, how do we find out whether our agent considers all the events in the partition to be equally probable? Here is an answer to this question, generalizing Ramsey's idea for how to find an event with subjective probability $\frac{1}{2}$. First, we find outcomes $b$ and $w$ such that our agent strictly prefers $b$ over $w$. Now, we check whether our agent is indifferent between a gamble which yields $b$ if $E_i$ happens and $w$ otherwise, and a gamble which yields $b$ if $E_j$ happens and $w$ otherwise, for all distinct events $E_i$ and $E_j$ in our partition. This means that our agent does not care whether the `good prize' is on the first event in the partition, the second event in the partition and so on. Now, we assume the following:
\begin{quote}
\textbf{Generalized Better Prize Condition.} If an agent strictly prefers $b$ to $w$ and is indifferent between the gambles $\{E_1, b ; \textrm{not-}E_1, w\}$ and $\{E_2, b; \textrm{not-}E_2, w\}$, then our agent thinks that $E_1$ and $E_2$ are equally probable: $p(E_1) = p(E_2)$.
\end{quote}
Again, this assumption is entailed by both EU maximization and REU maximization.\footnote{Here is the argument: Since $\{E_1, b ; \textrm{not-}E_1, w\} \sim  \{E_2, b; \textrm{not-}E_2, w \}$, we have $r(p(E_1)) = r(p(E_2))$.  Since $r$ is strictly increasing, and so injective, $p(E_1) = p(E_2)$.} Then, we use this assumption to infer that our agent must consider all events to be equally probable, that is $p(E_i) = p(E_j)$ for all events $E_i$ and $E_j$ in our partition. Again, this is quite intuitive: If our agent doesn't care whether the good prize is on event $E_1$ or $E_2$ or .... or $E_n$, our agent must consider all events in the partition to be equally probable, even if our agent is sensitive to risk.

Now that we have our fair lottery $\{E_1, ... ,E_n\}$, we also have an event with probability $\frac{k}{n}$ with $k \leq n$. This is simply the event that one of the first $k$ tickets wins: $\bigcup_{i=1}^k E_i = E_1 \cup ... \cup E_k$. Since all events in $\mathcal{P}$ are equiprobable, we have $p(E_i) = \frac{1}{n}$ for all $i \leq n$. Further, all events in our lottery are disjoint, so 
\begin{equation*}
p(\bigcup_{i=1}^k E_i) = \sum_{i=1}^k p(E_i) = \frac{k}{n}.
\end{equation*}
Therefore, we have found our desired event $E$ with probability $p(E) = \frac{k}{n} = a$.

At this point, we can use Ramsey's method to determine $r(a)$. First, we find outcomes $b,m,w$ which satisfy Ramsey's condition, so
\begin{enumerate}[i.]
\item our agent strictly prefers $b$ over $w$: $w \prec b$,
\item our agent is indifferent between getting $m$ for certain and a gamble which yields $b$ if $E$ happens and $w$ otherwise: $m \sim \{b, E; w, \textrm{not-}E\}$.
\end{enumerate}
Then, we infer that $r(p(E)) = \frac{u(m) - u(w)}{u(b) - u(w)}$. Therefore, we know that $r(a) =  \frac{u(m) - u(w)}{u(b) - u(w)}$. Given the assumption that we have already determined our agent's utility function, we can compute $r(a)$. This is the decision weight our agent ascribes to events with probability $a = \frac{k}{n}$. Thus, we have a method to find decision weights for events with any rational-valued probability. As explained earlier, this gives us enough information to determine our agents' risk function $r$ uniquely. Therefore, we can measure the risk function of a REU maximizer by their observable preferences among gambles.

I described a method to measure our agent's attitude towards risk. This method relies on `fair lotteries': partitions of our event space into $n$ equiprobable events. Two remarks are in order. First, this method is not an algorithm for finding events with any rational-valued probability. This is because I have merely described a method for checking when our agent considers a partition to be a fair lottery, and no general method for finding such partitions. However, Ramsey does not provide an algorithm for finding outcomes which satisfy Ramsey's conditions. Therefore, the lack of an algorithm does not make our method worse than Ramsey's original method. 

Second, our method assumes that there are partitions of our agent's event space into arbitrarily many equiprobable events. Note that this goes beyond Ramsey's original assumptions. However, partitions of our agent's event space into arbitrarily many equiprobable events have a natural interpretation: they correspond to fair lotteries with arbitrarily many tickets. To find such partitions, we just have to find lotteries with arbitrarily many tickets which our agent considers to be fair. I admit that assuming the existence of such lotteries is still a considerable idealization. But in the serious business of decision theory, we have to make some idealizations. The important point is that, when it comes to idealizations in decision theory, assuming the existence of fair lotteries is a pretty mild one.\footnote{Due to a result by \cite{WS1922}, we know that such partitions exist if our agent's subjective probability function is \emph{countably additive} and \emph{non-atomic}. Also see \cite[p. 36]{Billingsley1995}. Relatedly, note that the existence of fair lotteries (or something stronger) has been quite frequently assumed to get from a qualitative ordering of events by their comparative probability to a unique probability measure which represents this ranking. See, for example, \cite{deFinetti1937, Koopman1940} and in particular the discussion in \cite[p. 38-39]{Savage1954}.}

\subsection{From Risk Attitude to Subjective Probability}

I close by explaining two different ways in which we can leverage the method of fair lotteries to measure the subjective probabilities of an REU maximizer.

First observe that the method of fair lotteries already gives us a straightforward way to measure all \emph{rational-valued} subjective probabilities. If our agent ascribes some rational probability value $\frac{k}{n}$ to event $E$, we can discover this by observing that our agent considers the event $E$ to be equally probable as the event that one of the first $k$ tickets in a fair lottery with $n$ tickets wins. Note, however, that this doesn't quite work for irrational-valued subjective probabilities. If our agent ascribes an irrational probability value to event $E$, such as $\frac{1}{\pi}$, then $E$ is not equally probable to any fair lottery event. Yet we can still use the method of fair lotteries to find pairs of events which are more and less probable than $E$, and then `squeeze' these two events closer and closer to each other. We obtain the probability of $E$ as the limit point of this process. Thus, we can use the method of fair lotteries to measure all subjective probabilities of a REU maximizer, although in a less constructive way than Ramsey's original method. 

Second, once we have measured out our agent's attitude towards risk, when can use this information to recover subjective probabilities from decision weights. Suppose we want to find out which subjective probability our agent ascribes to event $E$. We find outcomes $b,m,w$ which satisfy Ramsey's condition. Like Ramsey, we have to assume that our agent's outcome space is sufficiently rich to contain such outcomes for any event $E$ whose probability we want to measure. Then we use Ramsey's method, together with the assumption that our agent maximizes REU, to infer that $r(p(E)) = \frac{u(m) - u(w)}{u(b) - u(w)}$. Now since we know the risk function $r$, we can disentagle subjective probabilities and attitudes towards risk: 
\begin{equation*}
p(E) = r^{-1}(r(p(E))) = r^{-1}\left(\frac{u(m) - u(w)}{u(b) - u(w)}\right).
\end{equation*} 
Thus, we have a method to measure the subjective probabilities of a REU maximizer. The catch is that we first have to determine their risk function, which will take infinitely many steps. For this reason, this second method is also less constructive than Ramsey's original method.

\section{Conclusion}

Ramsey shows how we can measure the subjective probabilities of an agent by their observable preferences among gambles, assuming that the agent is an EU maximizer and we have enough information about their utility function. In this paper, I show how to extend the spirit of Ramsey's method to a strictly wider class of agents: REU maximizers. In particular, I show how we can measure the risk attitudes of an agent by their observable preferences among gambles, assuming that the agent is an REU maximizer and we have enough information about their utility function. I further explain how we can leverage this method to measure subjective probabilities of a REU maximizer. Since EU theory is a special case of REU theory, our method also works for EU maximizers. Thus, we have a method for measuring both  risk attitudes and subjective probability which is strictly more general than Ramsey's original method.

In a nutshell, the upshot is this: When we are using decision theory as a tool of hermeneutics, a method of making sense of each other, moving to a more permissive decision theory allows us to make sense of more kinds of agents. Ramsey's original insight was that we can use decision theory to get from people's overt behavior to an interpretation of their mental states, their beliefs and desires. I hope to have demonstrated that we can extend this great insight beyond the narrow confines of orthodox expected utility theory. This doesn't show that being sensitive to risk is \emph{rational} -- but it shows that we can make sense of it. The scope of `decision theoretic hermeneutics' includes agents which are sensitive to risk.

Let me close with two open questions which I hope to address in future work. First, our method only works if the utility function of the agent is already known.  But we would also like measure beliefs and risk attitudes if we have \emph{no information} about an agent's utility function. Thus, it would be great to get rid of the assumption that the utility function is known. Second, our method only works if the agent's outcome space is sufficiently rich -- for any event $E$ whose probability we want to measure, we need to find outcomes which satisfy Ramsey's condition. However, it would be great to extend our method to agents who do not make such fine-grained distinctions between outcomes. Note that both of these limitations also apply to Ramsey's original method. Thus, we need to generalize Ramsey even further that I have done here.

\bigskip

\textbf{Acknowledgments.} Thanks to audiences at the 7th CSLI Workshop on Logic, Rationality \& Intelligent Interaction at Stanford in 2018 and the 2019 Formal Epistemology Workshop in Turin, where earlier versions of this material were presented. Special thanks to Lara Buchak, Mikayla Kelley, Krzysztof Mierzewski and Yifeng Ding for helpful comments and discussion.

\bibliographystyle{eptcs}
\bibliography{bib}

\begin{thebibliography}{10}
\providecommand{\bibitemdeclare}[2]{}
\providecommand{\surnamestart}{}
\providecommand{\surnameend}{}
\providecommand{\urlprefix}{Available at }
\providecommand{\url}[1]{\texttt{#1}}
\providecommand{\href}[2]{\texttt{#2}}
\providecommand{\urlalt}[2]{\href{#1}{#2}}
\providecommand{\doi}[1]{doi:\urlalt{http://dx.doi.org/#1}{#1}}
\providecommand{\bibinfo}[2]{#2}

\bibitemdeclare{book}{Abott2001}
\bibitem{Abott2001}
\bibinfo{author}{Stephen \surnamestart Abbott\surnameend}
  (\bibinfo{year}{2001}): \emph{\bibinfo{title}{Understanding Analysis}}.
\newblock \bibinfo{publisher}{Springer}, \doi{10.1007/978-1-4939-2712-8}.

\bibitemdeclare{article}{Allais1953}
\bibitem{Allais1953}
\bibinfo{author}{Maurice \surnamestart Allais\surnameend}
  (\bibinfo{year}{1953}): \emph{\bibinfo{title}{Le Comportement de l'Homme
  Rationnel devant le Risque: Critique des Postulats et Axiomes de l'Ecole
  Americaine}}.
\newblock {\sl \bibinfo{journal}{Econometrica}}
  \bibinfo{volume}{21}(\bibinfo{number}{4}), pp. \bibinfo{pages}{503--546},
  \doi{10.2307/1907921}.

\bibitemdeclare{book}{Bermudez2011}
\bibitem{Bermudez2011}
\bibinfo{author}{Jos{\'e}~Luis \surnamestart Berm{\'u}dez\surnameend}
  (\bibinfo{year}{2011}): \emph{\bibinfo{title}{Decision Theory and
  Rationality}}.
\newblock \bibinfo{publisher}{Oxford University Press},
  \doi{10.1093/acprof:oso/9780199548026.001.0001}.

\bibitemdeclare{book}{Billingsley1995}
\bibitem{Billingsley1995}
\bibinfo{author}{Patrick \surnamestart Billingsley\surnameend}
  (\bibinfo{year}{1995}): \emph{\bibinfo{title}{Probability and Measure}},
  \bibinfo{edition}{3} edition.
\newblock \bibinfo{series}{Wiley Series in Probability and Mathematical
  Statistic}, \bibinfo{publisher}{Wiley}.

\bibitemdeclare{article}{Bradley2004}
\bibitem{Bradley2004}
\bibinfo{author}{Richard \surnamestart Bradley\surnameend}
  (\bibinfo{year}{2004}): \emph{\bibinfo{title}{Ramsey's Representation
  Theorem}}.
\newblock {\sl \bibinfo{journal}{Dialectica}}
  \bibinfo{volume}{58}(\bibinfo{number}{4}), pp. \bibinfo{pages}{483--97},
  \doi{10.1111/j.1746-8361.2004.tb00320.x}.

\bibitemdeclare{book}{Buchak2013}
\bibitem{Buchak2013}
\bibinfo{author}{Lara \surnamestart Buchak\surnameend} (\bibinfo{year}{2013}):
  \emph{\bibinfo{title}{Risk and Rationality}}.
\newblock \bibinfo{publisher}{Oxford University Press},
  \doi{10.1093/acprof:oso/9780199672165.001.0001}.

\bibitemdeclare{incollection}{Buchak2016}
\bibitem{Buchak2016}
\bibinfo{author}{Lara \surnamestart Buchak\surnameend} (\bibinfo{year}{2016}):
  \emph{\bibinfo{title}{Decision Theory}}.
\newblock In \bibinfo{editor}{Christopher \surnamestart Hitchcock\surnameend}
  \& \bibinfo{editor}{Alan \surnamestart Hajek\surnameend}, editors: {\sl
  \bibinfo{booktitle}{Oxford Handbook of Probability and Philosophy}},
  \bibinfo{publisher}{Oxford University Press}, pp. \bibinfo{pages}{789--814},
  \doi{10.1093/oxfordhb/9780199607617.013.40}.

\bibitemdeclare{article}{Buchak2017}
\bibitem{Buchak2017}
\bibinfo{author}{Lara \surnamestart Buchak\surnameend} (\bibinfo{year}{2017}):
  \emph{\bibinfo{title}{Taking Risks Behind the Veil of Ignorance}}.
\newblock {\sl \bibinfo{journal}{Ethics}}
  \bibinfo{volume}{127}(\bibinfo{number}{3}), pp. \bibinfo{pages}{610--44},
  \doi{10.1086/690070}.

\bibitemdeclare{article}{Davidson1973}
\bibitem{Davidson1973}
\bibinfo{author}{Donald \surnamestart Davidson\surnameend}
  (\bibinfo{year}{1973}): \emph{\bibinfo{title}{Radical Interpretation}}.
\newblock {\sl \bibinfo{journal}{Dialectica}}
  \bibinfo{volume}{27}(\bibinfo{number}{1}), pp. \bibinfo{pages}{314--28},
  \doi{10.1111/j.1746-8361.1973.tb00623.x}.

\bibitemdeclare{article}{deFinetti1937}
\bibitem{deFinetti1937}
\bibinfo{author}{Bruno \surnamestart De~Finetti\surnameend}
  (\bibinfo{year}{1937}): \emph{\bibinfo{title}{La Pr\'evision: Ses Lois
  Logiques, Ses Sources Subjectives}}.
\newblock {\sl \bibinfo{journal}{Annales de l'Institut Henri Poincar\'e}}
  \bibinfo{volume}{17}, pp. \bibinfo{pages}{1--68}.

\bibitemdeclare{article}{Elliott2017}
\bibitem{Elliott2017}
\bibinfo{author}{Edward \surnamestart Elliott\surnameend}
  (\bibinfo{year}{2017}): \emph{\bibinfo{title}{Ramsey Without Ethical
  Neutrality: A New Representation Theorem}}.
\newblock {\sl \bibinfo{journal}{Mind}}
  \bibinfo{volume}{126}(\bibinfo{number}{501}), pp. \bibinfo{pages}{1--51},
  \doi{10.1093/mind/fzv180}.

\bibitemdeclare{article}{Fishburn1981}
\bibitem{Fishburn1981}
\bibinfo{author}{Peter~C. \surnamestart Fishburn\surnameend}
  (\bibinfo{year}{1981}): \emph{\bibinfo{title}{Subjective Expected Utility: A
  Review of Normative Theories}}.
\newblock {\sl \bibinfo{journal}{Theory and Decision}}
  \bibinfo{volume}{13}(\bibinfo{number}{2}), pp. \bibinfo{pages}{139--99},
  \doi{10.1007/BF00134215}.

\bibitemdeclare{book}{Jeffrey1990}
\bibitem{Jeffrey1990}
\bibinfo{author}{Richard \surnamestart Jeffrey\surnameend}
  (\bibinfo{year}{1990}): \emph{\bibinfo{title}{The Logic of Decision}},
  \bibinfo{edition}{2} edition.
\newblock \bibinfo{publisher}{University of Chicago Press},
  \doi{10.2307/2183328}.

\bibitemdeclare{article}{KW2003}
\bibitem{KW2003}
\bibinfo{author}{Veronika \surnamestart K\"obberling\surnameend} \&
  \bibinfo{author}{Peter~P. \surnamestart Wakker\surnameend}
  (\bibinfo{year}{2003}): \emph{\bibinfo{title}{Preference Foundations for
  Nonexpected Utility: A Generalized and Simplified Technique}}.
\newblock {\sl \bibinfo{journal}{Mathematics of Operations Research}}
  \bibinfo{volume}{28}(\bibinfo{number}{3}), pp. \bibinfo{pages}{395--423},
  \doi{10.1287/moor.28.3.395.16390}.

\bibitemdeclare{article}{Koopman1940}
\bibitem{Koopman1940}
\bibinfo{author}{Bernard~O. \surnamestart Koopman\surnameend}
  (\bibinfo{year}{1940}): \emph{\bibinfo{title}{The Bases of Probability}}.
\newblock {\sl \bibinfo{journal}{American Mathematical Society}}
  \bibinfo{volume}{46}, pp. \bibinfo{pages}{763--74},
  \doi{10.1090/S0002-9904-1940-07294-5}.

\bibitemdeclare{article}{Lewis1974}
\bibitem{Lewis1974}
\bibinfo{author}{David \surnamestart Lewis\surnameend} (\bibinfo{year}{1974}):
  \emph{\bibinfo{title}{Radical Interpretation}}.
\newblock {\sl \bibinfo{journal}{Synthese}}
  \bibinfo{volume}{27}(\bibinfo{number}{3-4}), pp. \bibinfo{pages}{331--44},
  \doi{10.1007/BF00484599}.

\bibitemdeclare{article}{Machina1987}
\bibitem{Machina1987}
\bibinfo{author}{Mark~J. \surnamestart Machina\surnameend}
  (\bibinfo{year}{1987}): \emph{\bibinfo{title}{Choice under Uncertainty:
  Problems Solved and Unsolved}}.
\newblock {\sl \bibinfo{journal}{Journal of Economic Perspectives}}
  \bibinfo{volume}{1}(\bibinfo{number}{1}), pp. \bibinfo{pages}{121--54},
  \doi{10.1257/jep.1.1.121}.

\bibitemdeclare{article}{machina_more_1992}
\bibitem{machina_more_1992}
\bibinfo{author}{Mark~J. \surnamestart Machina\surnameend} \&
  \bibinfo{author}{David \surnamestart Schmeidler\surnameend}
  (\bibinfo{year}{1992}): \emph{\bibinfo{title}{A {More} {Robust} {Definition}
  of {Subjective} {Probability}}}.
\newblock {\sl \bibinfo{journal}{Econometrica}}
  \bibinfo{volume}{60}(\bibinfo{number}{4}), pp. \bibinfo{pages}{745--80},
  \doi{10.2307/2951565}.

\bibitemdeclare{article}{Okasha2007}
\bibitem{Okasha2007}
\bibinfo{author}{Samir \surnamestart Okasha\surnameend} (\bibinfo{year}{2007}):
  \emph{\bibinfo{title}{Rational Choice, Risk Aversion, And Evolution}}.
\newblock {\sl \bibinfo{journal}{Journal of Philosophy}}
  \bibinfo{volume}{104}(\bibinfo{number}{5}), pp. \bibinfo{pages}{217--35},
  \doi{10.5840/jphil2007104523}.

\bibitemdeclare{article}{Oliver2009}
\bibitem{Oliver2009}
\bibinfo{author}{Adam \surnamestart Oliver\surnameend} (\bibinfo{year}{2003}):
  \emph{\bibinfo{title}{A quantitative and qualitative test of the Allais
  paradox using health outcomes}}.
\newblock {\sl \bibinfo{journal}{Journal of Economic Psychology}}
  \bibinfo{volume}{24}(\bibinfo{number}{1}), pp. \bibinfo{pages}{35--48},
  \doi{10.1016/S0167-4870(02)00153-8}.

\bibitemdeclare{book}{Parmigiani2009}
\bibitem{Parmigiani2009}
\bibinfo{author}{Giovanni \surnamestart Parmigiani\surnameend} \&
  \bibinfo{author}{Lurdes Y.~T. \surnamestart Inoue\surnameend}
  (\bibinfo{year}{2009}): \emph{\bibinfo{title}{Decision Theory: Principles and
  Approaches}}.
\newblock \bibinfo{series}{Wiley Series in Probability and Mathematical
  Statistic}, \bibinfo{publisher}{Wiley}, \doi{10.1002/9780470746684}.

\bibitemdeclare{article}{Quiggin1983}
\bibitem{Quiggin1983}
\bibinfo{author}{John \surnamestart Quiggin\surnameend} (\bibinfo{year}{1983}):
  \emph{\bibinfo{title}{A Theory of Anticipated Utility}}.
\newblock {\sl \bibinfo{journal}{Journal of Economic Behavior and
  Organization}} \bibinfo{volume}{3}(\bibinfo{number}{1}), pp.
  \bibinfo{pages}{323--43}, \doi{10.1016/0167-2681(82)90008-7}.

\bibitemdeclare{article}{Rabin2000}
\bibitem{Rabin2000}
\bibinfo{author}{Matthew \surnamestart Rabin\surnameend}
  (\bibinfo{year}{2000}): \emph{\bibinfo{title}{Risk aversion and
  expected-utility theory: A calibration theorem}}.
\newblock {\sl \bibinfo{journal}{Econometrica}}
  \bibinfo{volume}{68}(\bibinfo{number}{5}), pp. \bibinfo{pages}{1281--92},
  \doi{10.1111/1468-0262.00158}.

\bibitemdeclare{incollection}{Ramsey1926}
\bibitem{Ramsey1926}
\bibinfo{author}{F.~P. \surnamestart Ramsey\surnameend} (\bibinfo{year}{1926}):
  \emph{\bibinfo{title}{Truth and Probability}}.
\newblock In \bibinfo{editor}{R.B. \surnamestart Braithwaite\surnameend},
  editor: {\sl \bibinfo{booktitle}{The Foundations of Mathematics and other
  Logical Essays}}, \bibinfo{edition}{1999 electronic edition} edition,
  \bibinfo{publisher}{Harcourt}, pp. \bibinfo{pages}{156--98},
  \doi{10.4324/9781315887814}.

\bibitemdeclare{article}{Savage1971}
\bibitem{Savage1971}
\bibinfo{author}{Leonard \surnamestart Savage\surnameend}
  (\bibinfo{year}{1971}): \emph{\bibinfo{title}{Elicitation of Personal
  Probabilities and Expectations}}.
\newblock {\sl \bibinfo{journal}{Journal of the American Statistical
  Association}} \bibinfo{volume}{66}(\bibinfo{number}{336}), pp.
  \bibinfo{pages}{783--801}, \doi{10.2307/2284229}.

\bibitemdeclare{book}{Savage1954}
\bibitem{Savage1954}
\bibinfo{author}{Leonard~J. \surnamestart Savage\surnameend}
  (\bibinfo{year}{1954}): \emph{\bibinfo{title}{The Foundations of
  Statistics}}.
\newblock \bibinfo{publisher}{Wiley Publications in Statistics}.

\bibitemdeclare{article}{WS1922}
\bibitem{WS1922}
\bibinfo{author}{W.~\surnamestart Sierpi\'{n}ski\surnameend}
  (\bibinfo{year}{1922}): \emph{\bibinfo{title}{Sur les fonctions d'ensemble
  additives et continues}}.
\newblock {\sl \bibinfo{journal}{Fund. Math.}}
  \bibinfo{volume}{3}(\bibinfo{number}{1}), pp. \bibinfo{pages}{240--46},
  \doi{10.4064/fm-3-1-240-246}.

\end{thebibliography}

\end{document}